\documentclass[useAMS,usenatbib]{mn2e}
\usepackage{graphicx}
\usepackage{times}
\usepackage{amssymb}
\usepackage{natbib}
\def\gsim { \lower .75ex \hbox{$\sim$} \llap{\raise .27ex \hbox{$>$}} }
\def\lsim { \lower .75ex \hbox{$\sim$} \llap{\raise .27ex \hbox{$<$}} }
\begin{document}

\title{The Origin of Extended Disk Galaxies at z=2}

\author[Sales et al.]{
\parbox[t]{\textwidth}{
Laura V. Sales$^{1}$, 
Julio F. Navarro,$^{2}$, 
Joop Schaye$^{3}$, 
Claudio Dalla Vecchia$^{3}$,
Volker Springel$^{4}$, 
Marcel R. Haas$^{3}$,
Amina Helmi$^{1}$
}
\\
\\
$^{1}$ Kapteyn Astronomical Institute, P.O. Box 800, Groningen, The Netherlands\\
%\\
$^{2}$ Department of Physics and Astronomy, University of Victoria, Victoria, BC V8P 5C2,
Canada\\
$^{3}$ Leiden Observatory, Leiden University, PO Box 9513, 2300 RA Leiden, The Netherlands\\
$^{4}$Max Planck Institute for Astrophysics, Karl-Schwarzschild-Strasse 1, 85740 Garching, Germany\\
%%
%\\
}

\maketitle

\begin{abstract}
  Galaxy formation models typically assume that the size and rotation
  speed of galaxy disks are largely dictated by the mass,
  concentration, and spin of their surrounding dark matter
  haloes. Equally important, however, are the fraction of baryons in
  the halo that collect into the central galaxy, as well as the net
  angular momentum that they are able to retain during its assembly
  process.  We explore the latter using a set of four large
  cosmological N-body/gasdynamical simulations drawn from the OWLS
  (OverWhelmingly Large Simulations) project. These runs differ only
  in their implementation of feedback from supernovae.  We find that,
  when expressed as fractions of their virial values, galaxy mass and
  net angular momentum are tightly correlated.  Galaxy mass fractions,
  $m_d=M_{\rm gal}/M_{\rm vir}$, depend strongly on feedback, but only
  weakly on halo mass or spin over the halo mass range explored here
  ($M_{\rm vir}>10^{11}\,  h^{-1}M_\odot$).  The angular momentum of a
  galaxy, expressed in units of that of its surrounding halo,
  $j_d=J_{\rm gal}/J_{\rm vir}$, correlates with $m_d$ in a manner
  that is insensitive to feedback and that deviates strongly from the
  simple $j_d=m_d$ assumption often adopted in semi-analytic models of
  galaxy formation.  The $m_d$-$j_d$ correlation implies that, in a
  given halo, galaxy disk size is maximal when the central galaxy
  makes up a substantial fraction ($\sim 20$-$30\%$) of all baryons
  within the virial radius (i.e., $m_d\sim 0.03$-$0.05$). At $z=2$,
  such systems may host gaseous disks with radial scale lengths as
  large as those reported for star-forming disks by the SINS survey,
  even in moderately massive haloes of average spin. Extended disks at
  $z=2$ may thus signal the presence of systems where galaxy formation
  has been particularly efficient, rather than the existence of haloes
  with unusually high spin parameter.
\end{abstract}

\section{Introduction}
\label{sec:intro}

Galaxy disks are widely assumed to form as gas cools and flows
hydrodynamically to the center of dark matter haloes. There, baryons
settle into thin, rotationally-supported structures whose size and
rotation speed are determined by the detailed mass profile of the
system and the net angular momentum of the cooled baryons. In this
formation scenario, first laid out by \cite{Fall1980} and worked out
in detail by \citet[][hereafter MMW98]{Mo1998}, the tight correlations
observed to link the mass, size, and rotation speed of galaxy disks
are thought to reflect analogous correlations between the mass, size
and spin of their surrounding dark haloes \citep[see,
e.g.,][]{Navarro2000}.

For dark haloes, these scaling laws have been extensively studied
through cosmological N-body simulations, and are now fairly well
understood. Mass and radius scale so that galaxy haloes are all
systems of roughly similar overdensity that, when properly scaled,
follow a mass profile well approximated by a simple ``universal''
fitting formula \citep{Navarro1996,Navarro1997}. Similarly, the net
spin of dark haloes, when expressed in dimensionless form ($\lambda$),
appears on average to be independent of mass and/or redshift
\citep[see, e.g.,][]{Bett2007}.

One important and robust corollary of these results is that the
characteristic density of haloes should broadly track the mean density
of the universe and, therefore, increase rapidly with redshift. On
dimensional grounds, since circular velocity scales like $V\propto
\sqrt{GM/R} \propto \sqrt{G\rho R^2}$, an increase in density, $\rho$,
implies also an increase in the ratio $V/R$ with $z$. Therefore, if
the size and rotation speed of galaxy disks simply reflect the radius
and circular velocity of their surrounding haloes, the size of disks
of given rotation speed is expected to decrease steadily with
increasing $z$.

The recent discovery of a population of extended disk galaxies at
$z=2$ by the SINS survey has confounded this simple expectation
\citep{Genzel2006,Foerster2006}. In some cases, these galaxies have
rotation speeds comparable to those of $L_*$ disks today and are as
extended as their $z=0$ counterparts, a result that has led
\cite{Bouche2007} to argue that haloes with unusually large
spin---much greater than expected in the prevailing $\Lambda$CDM
paradigm---may be required to explain disks as large as observed in
the SINS survey.

We note, however, that disk size depends not only on the properties of
dark haloes, but also on the fraction of baryons that collapse to form
the disk, and on the net angular momentum that these baryons retain
during the disk assembly process. These parameters are difficult to
estimate, as they likely depend in a complex manner on the intricate
process of mass accretion, star formation, and the regulating effects
of energetic feedback that accompany the formation of a galaxy. A
further complication stems from the fact that SINS observations
measure the size of the star-forming gaseous disk through its
redshifted $H_{\alpha}$ emission. This may differ from the true
spatial extension of all the baryons in the disk, which is what
theoretical models predict most accurately.

We address these issues here using a set of large cosmological
N-body/gasdynamical simulations from the OWLS (OverWhelmingly Large
Simulations) project (Schaye et al, in preparation). In this Letter we
study the dependence of galaxy disk size on supernova feedback and its
consequence for the interpretation of $z=2$ galaxy sizes reported by
the SINS collaboration. A more general analysis of the relation
between galaxy mass, angular momentum, feedback, and accretion history
will be presented in a companion paper (Sales et al, in preparation).

\section{The Numerical Simulations}
\label{sec:sims}

%%%%%%%%%%%%%%%%%%%%%%%%%%
\begin{center}
\begin{figure}
\includegraphics[width=80mm]{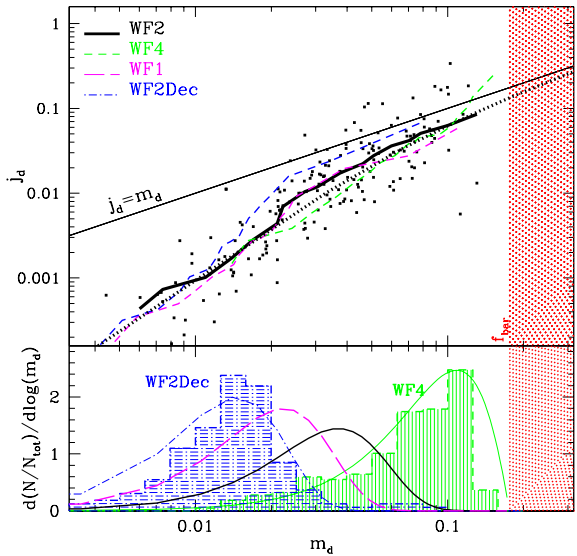}
\caption{{\it Top:} Galaxy mass ($m_d=M_{\rm gal}/M_{\rm vir}$) versus
  angular momentum ($j_d=J_{\rm gal}/J_{\rm vir}$) of simulated
  galaxies at $z=2$. Dots correspond to individual systems in the WF2
  run; the thick solid curve traces the median $j_d$ as a function of
  $m_d$. Other curves show the median $j_d$ for the other three
  runs. The $j_d$-$m_d$ relation departs significantly from the often
  assumed $j_d=m_d$ relation (top thin solid line).  Note that the
  $j_d$-$m_d$ relation is approximately the same in all runs and is
  well approximated by Eq. (1): $j_d=9.71\, m_d^2(1-\rm
  exp[-1/(9.71m_d)])$ (dotted line).  The shaded region on the right
  indicates where $m_d$ exceeds the universal baryon fraction, $f_{\rm
    bar}$. {\it Bottom:} Distribution of $m_d$ for each run. We show
  the histograms for WF4 and WF2Dec, our two extreme feedback models,
  as well as fits with Eq.~\ref{eq:md} for all four runs.}
\label{fig:fig1}
\end{figure}
\end{center}
%%%%%%%%%%%%%%%%%%%%%%%%%%

The OWLS project is a collection of over $60$ N-body/gasdynamical
cosmological simulations with varying numerical resolution and various
choices for the implementation of star formation, cooling, feedback,
and other sub-grid physics. All simulations use a substantially
modified version of the smoothed particle hydrodynamics code GADGET-3
\citep{Springel2005b}.

We select for this study four OWLS runs that differ from each other
only in the way energetic feedback from massive stars is
implemented. All these runs assume a standard $\Lambda$CDM cosmology
consistent with the WMAP-3 results ($\Omega_{\rm M}=0.238$,
$\Omega_{\rm CDM}=0.1962$, $\Omega_{\Lambda}=0.762$, $\Omega_{\rm
bar}=0.0418$, $h=0.73$, $\sigma_8=0.74$, $n=0.951$) and follow the
evolution of $512^3$ dark matter and $512^3$ gas particles in a $25 \,
h^{-1}$ Mpc box to $z=2$. The mass per baryonic particle is $\sim 1
\times 10^6 \, h^{-1} {\rm M}_\odot$ and $4.7$ times higher for the
dark matter component. The runs adopt a gravitational softening that
never exceeds $0.5\, h^{-1}$ kpc (physical).

Gas is allowed to cool radiatively using the element-by-element
implementation of Wiersma et al. (2009a). Gas with densities exceeding
$n_{\rm H}=0.1~{\rm cm}^{-3}$ is allowed to form stars at a
pressure-dependent rate consistent with the Schmidt-Kennicutt relation
\citep{Schaye2008}. The timed release of 11 chemical elements by
evolving stellar populations is implemented as described in Wiersma et
al. (2009b). \nocite{Wiersma2009a,Wiersma2009b} Feedback energy from
supernovae is incorporated assuming that a fixed fraction of the
available energy ($40\%$ of the total energy released by supernovae
for the adopted Chabrier IMF) is channeled into ``winds'' outflowing
from regions of active star formation. All four runs assume that the
total energy (per solar mass of stars formed) invested in the outflow
is the same, but differ in their numerical implementation, which is
controlled by two parameters: the wind velocity ($v_{\rm w}$) and the
mass loading ($\eta$) factor. The parameter $\eta$ specifies the
number of gas particles among which the feedback energy from a single
star particle is split, whereas $v_{\rm w}$ characterizes the outflow
velocity of particles in the wind (see \citealt{DallaVecchia2008} for
details).

For massive galaxies, the overall effect of feedback, as measured by
how effectively it regulates star formation and/or removes gas from
star-forming galaxies, increases with $v_{\rm w}$ for fixed $\eta \,
v_{\rm w}^2$. We refer to each of the 4 runs, in order of increasing
feedback efficiency, as: WF4 ($\eta=4$ and $v_{\rm w}=424$ km/s), WF2
($\eta=2$ and $v_{\rm w}=600$ km/s), WF1 ($\eta=1$ and $v_{\rm w}=848$
km/s) and WF2Dec. The latter is equivalent to WF2 but ``wind''
particles are temporarily decoupled from the hydrodynamical equations,
allowing them to freely escape the interstellar medium.

We focus on all simulated galaxies at the centers of haloes in the
virial mass range $10^{11}<M_{\rm vir}/(h^{-1} {\rm M}_\odot)<3\times
10^{12}$.  We define the virial radius, $r_{\rm vir}$, of a system as
that of a sphere enclosing a mean overdensity 178 times the critical
value at that redshift. All ``virial'' quantities are measured at or
within that radius.  Our galaxy sample contains $\sim 170$ objects
with between $50,000$ and $500,000$ particles within the virial
radius. The center is given by the position of the most-bound particle
and the center of mass velocity corresponds to the average velocity of
the 1000 innermost particles.  Central ``galaxies'' are identified in
each halo with the baryonic component of the system defined to be
contained within $r_{\rm gal}=0.1 \, r_{\rm vir}$ from the halo
center. We have explicitly checked that this definition includes all
of the stars and cold, star-forming gas obviously associated with the
central object. Since the four simulations have identical initial
conditions, any difference in the galaxy population can be traced
directly to the different feedback implementation in each run.

\section{Results}
\subsection{Galaxy mass and angular momentum}

The baryonic mass of central galaxies depends strongly on the details
of the feedback implementation. This is shown in the bottom panel of
Fig.~\ref{fig:fig1}, where we plot the distribution of galaxy masses
(gas and stars within $r_{\rm gal}$), expressed in units of the virial
mass of the surrounding halo, $m_d \equiv M_{\rm gal}/M_{\rm vir}$,
for our four OWLS runs at $z=2$. When feedback effects are weak, such
as in WF4, most of the baryons flow unimpeded to the center of the
halo. As a result, $m_d$ is on average quite large and approaches in
some systems the theoretical maximum posed by the universal baryon
fraction, $f_{\rm bar}=\Omega_{\rm bar}/\Omega_{\rm m}=0.17$. At the
other extreme, when feedback is strong, as in WF2Dec, a large fraction
of baryons are pushed out of the central galaxy by feedback-driven
winds. This leads to low values of $m_d$ ($\lsim \, 0.02$), implying
that fewer than $\sim 8\%$ of all available baryons are retained by
the central galaxy in a typical halo. Intermediate feedback choices
(as in WF1 and WF2) lead to $m_d$ distributions that straddle those
two extremes. In all cases, however, the $m_d$ distribution is quite
broad and insensitive to halo spin and mass, at least over the range
of masses included in our sample.

One surprise, given the sensitive dependence of $m_d$ on feedback
discussed above, is the fact that the net angular momentum of the
central galaxy, when expressed in units of the virial value ($j_d
\equiv J_{\rm gal}/J_{\rm vir}$), correlates with $m_d$, in a manner
that is approximately independent of feedback. This is shown in the
top panel of Fig.~\ref{fig:fig1}, where the thick line track the
median $j_d$ as a function of $m_d$. Points show individual values
only for WF2, others are omitted for clarity. (The same relation is
found in a simulation identical to WF2 but with $8\times$ lower mass
resolution, indicating that this result is not compromised by
numerical resolution.) All four runs follow the same $m_d$-$j_d$
trend, which is well approximated by the relation
\begin{equation}
j_d=9.71 \, m_d^2\, (1-\exp[-1/(9.71\,m_d)])
\label{eq:jdmd}
\end{equation}
where the last factor ensures that $j_d$ asymptotically approaches
$m_d$ for large $m_d$ values (see dotted line in Figure
\ref{fig:fig1}). Our simulations only sample the region $ 0.005 ~< m_d
~< 0.1$.  Care should therefore be taken when extrapolating
eq.~(\ref{eq:jdmd}) beyond these limits. The scatter about this
relation is not negligible, with an rms of order $\sigma_{\log
j_d}\sim 0.33$ about the median trend. Note that this relation implies
that the specific angular momentum of a galaxy, $j_{\rm gal} = J_{\rm
gal}/M_{\rm gal}$, is quite different from the specific angular
momentum of its surrounding halo, $j_{\rm vir}=J_{\rm vir}/M_{\rm
vir}$, unlike what is often assumed in semi-analytic models of galaxy
formation ($j_{\rm gal}=j_{\rm vir}$ is equivalent to setting
$j_d=m_d$). Indeed, $j_{\rm gal}$ and $j_{\rm vir}$ are very poorly
correlated in our simulations \citep[see also
][]{vandenBosch2002,Dutton2009}.  

\subsection{Galaxy sizes} 

%%%%%%%%%%%%%%%%%%%%%%%%%%
\begin{center}
\begin{figure}
\includegraphics[width=80mm]{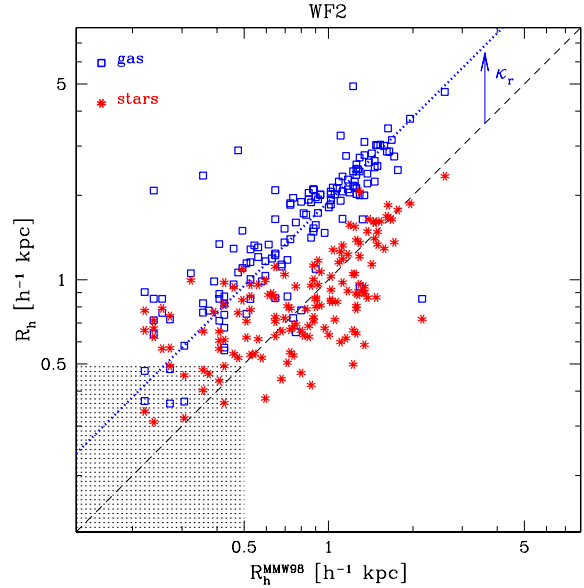}
\caption{Half mass radii predicted by the MMW98 formalism ($x$-axis)
  versus the half-mass radii of the stellar (starred symbols) and
  gaseous (open squares) components of simulated galaxies at $z=2$.
  Note the roughly constant offset in size between gas and MMW98
  predictions.}
\label{fig:fig2}
\end{figure}
\end{center}
%%%%%%%%%%%%%%%%%%%%%%%%%%

According to the MMW98 formalism, once $j_d$ and $m_d$ are specified,
the size of a galactic disk may be predicted in a halo of known virial
radius $r_{\rm vir}$, concentration $c$, and spin parameter
$\lambda$. The model assumes that baryons settle onto a thin,
rotationally supported disk with an exponential radial density
profile. The scale radius $R_d$ of such a disk is given by
\begin{equation} 
R_d={1\over {(2\,f_c)}^{1/2}}\, {\lambda\, j_d\over m_d}\, f_R(j_d,m_d,c,\lambda) \, r_{\rm vir},
\label{eq:rd}
\end{equation}
where $f_R$ and $f_c$ are known functions that account for the effect
of halo concentration as well as its adiabatic ``contraction'' in
response to the assembly of the galaxy. 

It is instructive to see how this simple prediction fares in
comparison with simulated galaxies. This is shown in
Fig.~\ref{fig:fig2}, where we compare half-mass radii measured for
galaxies identified in the WF2 run with those computed using
Eq.~(\ref{eq:rd}). We use half-mass radii instead of exponential scale
lengths because it is a more robust size estimate, especially when
profiles deviate from simple exponentials, typically as a result of
the presence of a spheroid. Starred symbols correspond to the stellar
component and open squares to the gaseous (star forming) disk
component. Half-mass radii are computed in projection, after using the
angular momentum of the galaxy to turn the system ``face-on''. The
shaded area indicates radii smaller than the gravitational softening.

On scales exceeding the softening radius, the agreement between the
size of the stellar component and the MMW98 prediction is reasonable,
especially considering the many simplifying assumptions in the
model. For example, MMW98 assume that all of the mass is in an
exponential disk, whereas many of the simulated galaxies show
prominent spheroidal components.

Encouragingly, the gaseous disks show a much better correlation with
MMW98 model, and seem to differ from $R_h^{\rm MMW98}$ by an 
approximately constant factor $\kappa_r \approx 1.8$. This implies 
that the predictions of Eq.~(\ref{eq:rd}) may be simply corrected by 
$\kappa_r$ to provide reasonably accurate estimates of the size of 
{\it gaseous} disks such as those observed in the SINS survey.

\subsection{Disk size and rotation speed}

The MMW98 formalism predicts not only the size of a disk but also its
rotation speed, $V_{\rm rot}$, once the halo parameters ($r_{\rm
vir}$, $c$, and $\lambda$) as well as the parameters characterizing
the galaxy ($m_d$ and $j_d$) are specified. As discussed in the
previous section, although $m_d$ and $j_d$ are tightly linked, the
$m_d$ distribution is very broad, so it is important to explore how
much disk sizes may vary solely as a result of variations in $m_d$.

\begin{center}
\begin{figure}
\includegraphics[width=80mm]{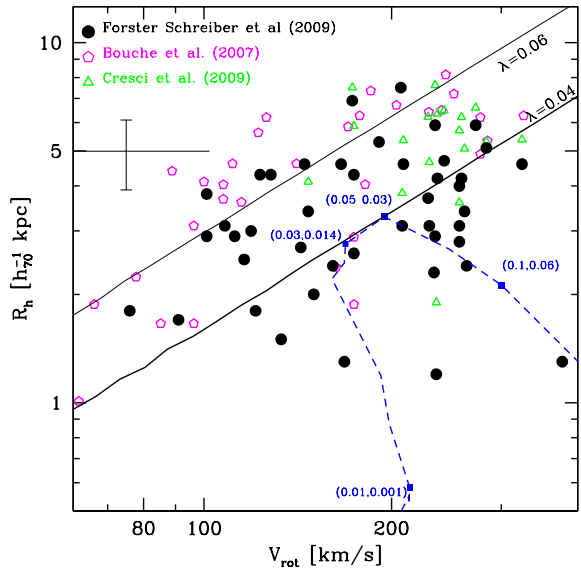}
\caption{The radius-velocity plane of observed high redshift galaxies
  from the SINS survey: Bouch\'e et al. (2007, magenta pentagons),
  Cresci et al. (2009, green triangles) and F{\"o}rster Schreiber et
  al. (2009, dark circles) (typical errors are indicated on the top left).
  These are compared with the half-mass
  radii of gaseous disks, computed using Eq.~(\ref{eq:rd}) and
  corrected by the $\kappa_r=1.8$ factor.  The dashed line corresponds
  to a disk in a halo with $V_{\rm vir}=150$ km$/$s, spin parameter
  $\lambda=0.04$, and concentration $c=4$; with a $j_d-m_d$ relation
  as given by the thick solid curve in Fig~\ref{fig:fig1}. Pairs of
  $(m_d,j_d)$ values are quoted to help interpretation. The solid
  curves outline the size-velocity relation expected for disks in
  haloes with two different spin parameters ($\lambda=0.04$ and
  $0.06$) and $m_d=0.05$. }
  \label{fig:fig3}
\end{figure}
\end{center}

We illustrate this in Fig.~\ref{fig:fig3} for the case of a halo with
virial velocity $V_{\rm vir}=150~{\rm km}/{\rm s}$, $c=4$ \citep[the
average concentration for that mass at $z=2$;][]{Gao2008,Duffy2008},
and $\lambda=0.04$ \citep[roughly the median spin parameter of $\Lambda$CDM
haloes; see, e.g.,][]{Bett2007}. We use the MMW98 formalism to compute
the size and rotation speed of a {\it gaseous} disk galaxy in this
halo as a function of $m_d$, assuming that $j_d(m_d)$ follows the
median relation for run WF2 (thick solid line in Fig.~\ref{fig:fig1})
and including the $\kappa_r$ correction.  The result is shown by the
dashed curve in Fig.~\ref{fig:fig3}, which spans a very wide range,
from $m_d=0.01$ at the bottom left endpoint to $m_d=0.1$ at the upper
right.

A few points are highlighted by this exercise. The first is the very
wide range in disk size expected from a broad $m_d$ distribution: the
disk size changes by a factor of roughly $\sim 7$, from $\sim 600$ pc
to almost $\sim 4$ kpc (physical), when $m_d$ increases from $0.01$ to
$0.05$. This implies that the size of galactic disks is as sensitive
to $m_d$ as to the spin parameter.

The second point is that disk size is maximized for $m_d\approx 0.05$,
and declines when $m_d$ increases further. This is because when $m_d$
exceeds a certain value, the disk becomes so massive that it begins to
significantly steepen the central potential, which in turn implies
that higher rotation speeds are required to achieve centrifugal
equilibrium. As a result, disks rotate faster but actually become
smaller, leading the dashed curve in Fig.~\ref{fig:fig3} to decline
for $m_d>0.05$.  Accurate predictions of the size of galactic disks
clearly require that $m_d$ and $j_d$ be very tightly constrained.

The pentagons, triangles, and circles in Fig.~\ref{fig:fig3} show SINS
data from the compilations of \cite{Bouche2007}, \cite{Cresci2009},
and \cite{Forster2009}.  Galaxy radii from \citet{Bouche2007} are
$H_{\alpha}$ half-light radii (taken from their Fig. 1) whereas for
Cresci et al (2009) the radii plotted are the ``disk scale lengths''
derived from their dynamical modeling (taken from their Table
2). Radii for the F{\"o}rster Schreiber et al galaxies are also
$H_{\alpha}$ half-light radii, as given in their Fig.18.

Although we show predicted half-mass gas radii in Fig.~\ref{fig:fig3},
we have also estimated in our simulated galaxies $H_{\alpha}$
half-light radii by weighting each gas particle by its star formation
rate and by applying an extinction correction based on the metal column
through the star-forming gas. This procedure gives results
which are similar to the simple half-mass radii of the gaseous disks
measured directly from the simulations, suggesting that the comparison
of SINS disk sizes with the half-mass gas radii of simulated galaxies
is fair.

The dashed-curve in Fig.~\ref{fig:fig3} thus makes clear that even a
moderately massive halo of average concentration and spin may host a
disk as large as some of those reported in the SINS survey ($\approx
3$-$4$ kpc if $m_d\approx 0.03$-$0.05$). Of course, larger disks may
also exist; for example, in more massive haloes, or in haloes with
larger-than-average spin, or in systems that scatter above the mean
$j_d$-$m_d$ relation shown in Fig.~\ref{fig:fig1}. For example, the
curves in Fig.~\ref{fig:fig3} labelled $\lambda=0.04$ and
$\lambda=0.06$ indicate where disks would lie if $m_d$ is fixed at
$0.05$ and the halo mass is varied. Disks as large as $\sim 10$ kpc
are clearly allowed for the right combination of $\lambda$ and $m_d$.

We conclude that the presence of some extended gaseous disks at $z=2$
is not necessarily in conflict with the $\Lambda$CDM scenario
\citep[see also ][]{Dekel2009b}. However, without strong and tight
constraints on the $m_d$ distribution at $z=2$ it is quite difficult
to assess whether the existence of extended disks in the SINS survey
presents a real challenge to the $\Lambda$CDM paradigm. Constraining
the $m_d$ distribution is, unfortunately, a non trivial task where
direct simulations offer little guidance, given the strong sensitivity
of $m_d$ to the uncertain implementation of feedback highlighted
above.

On the other hand, semi-analytic models suggest that a fairly low
galaxy formation efficiency is required in order to account for the
galaxy stellar mass function and its evolution with
redshift. \cite{Conroy2009}, for example, argue that fewer than $\sim
10\%$ of baryons should be transformed into stars in galaxy-sized
haloes. This implies $m_d \, \lsim \, 0.017$. Should this constraint
hold true, it would imply that very few systems may, on average, reach
the $m_d\approx 0.05$ needed to maximize their disk size.  Figure
1 shows that many of our galaxies are more massive relative to their
haloes than inferred by Conroy \& Wechsler (2009), particularly if the
SN feedback is weak. Unless star formation is quenched in these
objects, these simulations would therefore overestimate the abundance of
massive galaxies if they were continued to lower redshifts.

%%%%%%%%%%%%%%%%%%%%%%%%%%
%%%%%%%%%%%%%%%%%%%%%%%%%%

We can make some progress by assuming that the $m_d$ distribution,
$dn/dm_d$, is independent of halo mass and spin (which approximately
holds over the mass range considered here) and by parametrizing it
with a simple function,
\begin{equation}
dn/dx=A {x\over \sigma_x^2} [e^{-((x-1)/\sigma_x)^2}-1],
\label{eq:md}
\end{equation}
where $x=m_d/f_{\rm bar}$ and $A$ is a normalization factor.  We note,
however, that $m_d$ does depend on mass when extending this analysis
to lower mass halos. Eq.~(\ref{eq:md}) provides an adequate fit to the
$m_d$ distribution in all our simulations, as shown by the thin lines
in the bottom panel of Fig.~\ref{fig:fig1}. The single parameter
$\sigma_x$ fully specifies the distribution; the location of the peak
varies monotonically with $\sigma_x$, shifting from $m_d\approx 0.11$
($\sigma_x=0.76$) for WF4 to $m_d \approx 0.015$ ($\sigma_x=0.01$) for
WF2Dec.

Adopting Eq.~(\ref{eq:md}) to describe the $m_d$ distribution, and
Eq.~(\ref{eq:jdmd}) for the $j_d(m_d)$ dependence (including scatter,
$\sigma_{\log j_d} = 0.33$), we can compute the abundance of extended
disks expected in the $\Lambda$CDM scenario by populating the dark
haloes in the simulations with disks of size consistent with these
distributions. For each halo we assign, at random, a concentration
taken from a log-normal distribution centered on $c=4$ \citep{Gao2008}
with width $\sigma_{\log c}=0.1$ \citep{Neto2007}, and a spin
parameter drawn from the actual $\lambda$ distribution of the
sample. The latter has median $0.028$ and dispersion $\sigma_{\rm log
\lambda}=0.25$, roughly consistent with Bett et al (2007).

\begin{center}
\begin{figure}
\includegraphics[width=74mm]{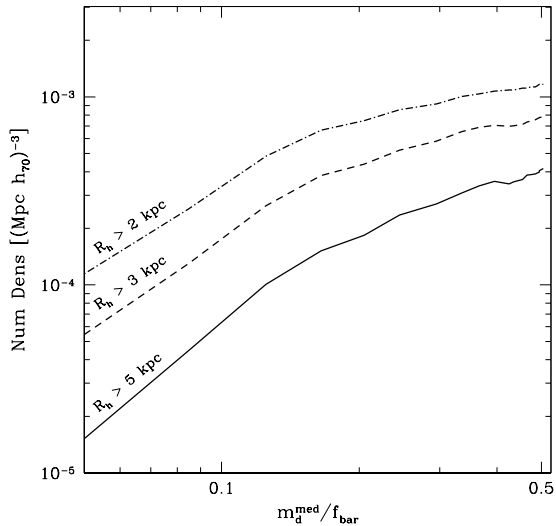}
\caption{Number density of extended gaseous disks expected in the
  $\Lambda$CDM scenario at $z=2$. The different curves show, as a
  function of the median $m_d$ (of a distribution given by
  Eq.~(\ref{eq:md})), the number of gaseous disks with $V_{\rm
  rot}>100$ km$/$s and half-mass radii exceeding $2$, $3$, and $5$
  $h_{70}^{-1} \rm kpc$.
\label{fig:fig4}}
\end{figure}
\end{center}

Fig.~\ref{fig:fig4} shows, as a function of $m_d^{\rm med}/f_{\rm
bar}$ (median of the $dn/dx$ distribution), the number density of
gaseous disks with $V_{\rm rot}>100$ km$/$s and half-mass radii
exceeding $2$, $3$, and $5 \, h_{70}^{-1}$ kpc (physical) at
$z=2$. For $m_d^{\rm med} \sim 0.1 f_{\rm bar}$, which corresponds to
the case where $\sim 10\%$ of baryons in a galaxy-sized halo have been
assembled into galaxies, we expect of order $\sim 1$ extended disk per
$10^4$ cubic comoving $h_{70}^{-1}$ Mpc. Even for a $m_d^{\rm
med}/f_{\rm bar}$ as low as $0.05$, one would still expect of order
one such system per $5 \times 10^5$ cubic $h_{70}^{-1}$ Mpc. Comparing
these numbers to the abundance of disks at $z=2$ from the SINS survey
would help to confirm whether the presence of extended disks is indeed
in conflict with $\Lambda$CDM.

%%%%%%%%%%%%%%%%%%%%%%%%%%
%%%%%%%%%%%%%%%%%%%%%%%%%%

\section{Conclusions}
\label{sec:concl}

We have analyzed the baryonic mass and angular momentum content of
galaxies identified in four different runs selected from the OWLS set
of cosmological N-body/gasdynamical simulations. We focus on galaxies
forming at the centers of haloes with virial mass between $10^{11}$
and $3\times 10^{12} \, h^{-1} {\rm M}_\odot$ at $z=2$. The four runs
analyzed differ only in the way the supernova feedback is implemented
numerically in the simulations.  Our main conclusions may be
summarized as follows.

\begin{itemize}

\item The efficiency of feedback, as measured by the fraction of
  baryons that assemble into a central galaxy in each halo, varies
  strongly in our four simulations. In the least efficient feedback
  run, WF4, typically about $50\%$ of the baryons in a halo get
  assembled into the central galaxy. At the other extreme, in run
  WF2Dec, where feedback efficiency is highest, central galaxies are
  able to retain less that $10\%$ of all available baryons.

\item Despite these differences in feedback efficiency, galaxies
identified in all four simulations follow the same tight relation
between mass (expressed in units of the halo virial mass, $m_d$) and
angular momentum (also expressed as a fraction of the virial value,
$j_d$). This relation deviates strongly from the simple $j_d$=$m_d$
assumption often used in semi-analytic models of galaxy formation and
is well approximated by $j_d=9.71\, m_d^2\,(1-\rm exp[-1/(9.71m_d)])$,
with rms dispersion $\sigma_{\log j_d} \sim 0.33$.

\item The galaxy mass fraction parameter, $m_d$, shows a broad
  distribution, with a mean and dispersion that are insensitive to
  halo spin and mass (over the range spanned by our sample). Together
  with the tight $j_d$-$m_d$ relation, this implies that the size of
  galactic disks will vary widely, even for galaxies formed in haloes
  of similar mass, concentration, and spin. The size of galactic disks
  is maximized when $m_d\approx 0.03$-$0.05$. Such disks can easily
  match the observed size of gaseous disks in the SINS survey, even in
  moderately massive haloes of average spin.

\item The abundance of extended disk systems expected at $z=2$ for
  $\Lambda$CDM depends critically on the $m_d$ distribution. Assuming
  that the form of this distribution may be approximated by
  Eq.~(\ref{eq:md}), we have computed the expected abundance of
  gaseous disks with half-mass radii exceeding $R_h=2$, $3$, and $5\,
  h_{70}^{-1}$ kpc (physical). If the average $m_d$ is such that $\sim
  10\%$ of baryons collect into galaxies, we would expect of order
  $\sim 1$ system with $R_h>3$ kpc in a $10^4$ Mpc$^3$ (comoving)
  volume. This predicted abundance could be compared with
  observational estimates, once they become available.

\end {itemize}

We conclude that galaxy formation efficiency, through the tight
$j_d$-$m_d$ relation found in our simulations, plays a crucial role in
setting the size and rotation speed of galaxy disks. This should be
taken into account carefully in galaxy formation models. Extended
disks are expected to exist at $z=2$ in the $\Lambda$CDM scenario, but
in small numbers. The large disks found by the SINS survey should not
be ``typical'' galaxies at $z=2$, but rather some of the largest. The
selection of SINS galaxies favors large disks in order to enhance the
probability of obtaining resolved velocity maps, so cannot be
considered ``typical' either. It seems likely that a number of large
disks at $z=2$ could be accommodated in the $\Lambda$CDM scenario
without resorting to unexpectedly high halo spins or other unusual
formation mechanism.

\vspace*{+0.5cm} We acknowledge the anonymous referee for useful
suggestions and comments that helped to improve the manuscript. We also
thank Natascha F{\"o}rster Schreiber for providing SINS data in
electronic format.  This work was supported by NWO, NOVA, NSF grant
PHY05-51164, and a Marie Curie Excellence Grant.

\bibliography{master}

\end{document}